\documentclass[amsmath,amssymb,amsbsy,prl,twocolumn,superscriptaddress,showpacs]{revtex4}
\usepackage{amsfonts}
\usepackage{amsmath}
\usepackage{amssymb}
\usepackage{bm}
\usepackage[usenames, dvipsnames]{color}
\usepackage{dcolumn}
\usepackage{epsfig}
\usepackage{eucal}
\usepackage{graphicx}
\usepackage[breaklinks,colorlinks = true,linkcolor = red,urlcolor=cyan,citecolor=red]{hyperref}
\usepackage{mathrsfs}
\usepackage{subfigure}
\usepackage{textcomp}
\usepackage{times}
\usepackage{xy}
\usepackage{color}

\begin{document}
\title{Understanding the Curious Magnetic State of Sr$_{3}$OsO$_{6}$}
\author{Shreya Das}
\affiliation{Department of Condensed Matter Physics and Material Sciences, S. N. Bose National Centre for Basic Sciences, JD Block, Sector III, Salt Lake, Kolkata, West Bengal 700106, India.}
\author{Anita Halder}
\affiliation{Department of Condensed Matter Physics and Material Sciences, S. N. Bose National Centre for Basic Sciences, JD Block, Sector III, Salt Lake, Kolkata, West Bengal 700106, India.}
\author{Atasi Chakraborty}
\affiliation{School of Physical Science, Indian Association for the Cultivation of Science, 2A \& 2B, Raja Subodh Chandra Mallick Rd, Jadavpur, Kolkata, West Bengal 700032, India.}
\author{Indra Dasgupta}
\affiliation{School of Physical Science, Indian Association for the Cultivation of Science, 2A \& 2B, Raja Subodh Chandra Mallick Rd, Jadavpur, Kolkata, West Bengal 700032, India.}
\author{Tanusri Saha-Dasgupta}
\email{t.sahadasgupta@gmail.com}
\affiliation{Department of Condensed Matter Physics and Material Sciences, S. N. Bose National Centre for Basic Sciences, JD Block, Sector III, Salt Lake, Kolkata, West Bengal 700106, India.}

\begin{abstract}
  Motivated by the recent report on high T$_c$ ferromagnetic insulating state of single transition metal containing double perovskite compound, Sr$_3$OsO$_6$ (Wakabayashi et. al., Nature Commun {\bf 10} 535, 2019), we study this curious behavior by employing first-principles calculations in conjunction with exact diagonalization of full $t_{2g}$ multiplet problem of two Os sites. Our analysis highlights the fact that stabilization of Sr$_3$OsO$_6$ in the cubic phase in epitaxially grown thin film is the key to both ferromagnetic correlation and high temperature scale associated to it. This also provides a natural explanation for the sister compound, Ca$_3$OsO$_6$ to exhibit low T$_N$ antiferromagnetism in its monoclinic structure. Further the insulating property is found to be driven by opening of Mott gap in the half filled spin-orbit coupled $j=3/2$ manifold of $d^{2}$ Os. We point out that Sr$_2$CaOsO$_6$ which naturally forms in the cubic phase would be worthwhile to explore as a future candidate to exhibit high T$_c$ ferromagnetic insulating state in bulk form. 
\end{abstract}  
\maketitle

\section{Introduction}

Ordered double perovskite compounds derived out of ABO$_3$ perovskites (A = alkaline earth/rare earth, B = transition metal)  with half of the B sites substituted by B$^{'}$, and rock-salt ordering of B-B$^{'}$, extending the formula to A$_2$BB$^{'}$O$_6$ have attracted significant attention primarily due to their intriguing electronic and magnetic properties.\cite{DDreview, Vasala, tsd} This includes above room temperature half metallic behaviour in compounds like Sr$_2$FeMoO$_6$ with transition temperature (T$_c$) of 420 K,\cite{tomioka, sfmo-p} Sr$_2$CrMoO$_6$ (T$_c$ = 420 K),\cite{claudia, anita} Sr$_2$CrWO$_6$ (T$_c$ = 458 K),\cite{CrW-Tc,hena} Sr$_2$CrReO$_6$ (T$_c$ = 620 K),\cite{CrRe-Tc} \textcolor{black}{ferrimagnetism in Ca$_2$MnOs$_6$,\cite{camnos} and Dirac-Mott insulating state in Ba$_2$NiOsO$_6$.\cite{banios}}  While all the above examples include two transition metals (TMs)
at B and B$^{'}$ sites, double perovskites with a single transition metal have been also synthesized. Most of these single TM containing double perovskites like Ba$_2$CaOsO$_6$,\cite{ba2caoso6} Sr$_2$MgOsO$_6$,\cite{sr2mgoso61, sr2mgoso62} Ca$_2$MgOsO$_6$\cite{sr2mgoso61} or Sr$_2$YReO$_6$ \cite{sr2yreo6} are either antiferromagnets or exhibit spin glass like behavior. The report of Sr$_3$OsO$_6$ \cite{Os-nat} double perovskite, containing a single {\it 5d} TM in B-B$^{'}$ sublattice, exhibiting ferromagnetic (FM) insulating state at a temperature $\approx$ 1060 K is thus unexpected and counter-intuitive. It is surprising on several counts. Firstly the TM-TM magnetic interaction is expected to be hindered by the
presence of a non magnetic Sr ion at B$^{'}$ site in the B-O-B$^{'}$-O-B super-exchange path. Secondly the high T$_c$ is expected to be a property of {\it 3d} TM containing compounds, rather than {\it 5d} TM containing compounds. Thirdly, most known ferromagnets are metals while insulating properties are attributed to antiferromagnets. Few known examples of ferromagnetic insulators are low T$_c$ materials {\it e.g.} EuO (77 K),\cite{euo} CdCr$_2$S$_4$ (90 K),\cite{cdcr2s4} SeCuO$_3$ (25 K)\cite{secuo} with the exception of La$_2$NiMnO$_6$,\cite{lanimno6} another FM insulating double perovskite with a T$_c$ of 280 K, which is still about a factor of 3-4 smaller compared to that reported for Sr$_3$OsO$_6$.

The situation becomes further puzzling by the fact that replacement of Sr by Ca makes the compound antiferromagnetic (AFM) with low transition temperature ($\approx$ 50 K).\cite{ca3oso6} It is curious to note while the crystal structure of the reported\cite{Os-nat} high T$_c$ ferromagnetic insulating compound Sr$_3$OsO$_6$ is cubic, that of Ca$_3$OsO$_6$ is monoclinic, prompting the role of crystal structure in influencing the magnetic behavior of these compounds. Os being a {\it 5d} element, the role of spin-orbit coupling ($\lambda$) becomes important and adds another dimension to the problem in addition to onsite Coulomb repulsion $U$ and Hund's coupling parameter $J_H$ within the multi orbital framework of Os $t_{2g}$'s. Understandably the interplay of crystal geometry, spin orbit coupling, Coulomb repulsion and Hund's coupling governs the exotic state of Sr$_3$OsO$_6$.

In the following we unravel this interplay by employing first-principles density functional theory calculations, constructing low energy model
Hamiltonian in first-principles derived Wannier basis and performing exact diagonalization calculation of a two Os site full multiplet problem with realistic parameters. Our analysis reveals that ferromagnetism in cubic symmetry
gets stabilized for a large range of $U$ and $J_H$, assisted by appreciably large nearest neighbor Os-Os hopping across the face of the cube of the double perovskite structure. Suppression of this hopping in distorted monoclinic symmetry destabilizes ferromagnetism and instead stabilizes antiferromagnetism. The stabilized ferromagnetic
state in cubic symmetry gives rise to an insulating solution through formation of a spin-orbit coupled Mott state of $d^{2}$, $j=3/2$ Os $t_{2g}$'s.
The reported\cite{Os-nat} cubic symmetry of Sr$_3$OsO$_6$, crucial for stabilization of ferromagnetic state, turns out to be not the natural choice, rather formed
in epitaxially grown thin film geometry. We propose Sr$_2$CaOsO$_6$ \cite{sr2caoso62} to be an alternative natural candidate which should
exhibit high T$_c$ FM insulating state even in the bulk form.

\section{Computational Details}

The first-principles density functional theory (DFT) calculations are carried out for realistic description of the problem. For the purpose of calculation, two different basis sets, namely
the plane wave basis\cite{plane} and muffin-tin orbital basis\cite{mto} are used. The agreement between the two sets of calculations
are verified in terms of calculated density of states and band structures.

The structural optimization as well as the influence of spin-orbit coupling (SOC) and onsite correlation $U$ on the electronic structure are investigated using plane-wave pseudo-potential method implemented within the Vienna Ab initio simulation package (VASP).\cite{VASP} In the plane-wave calculations the wave functions in the plane-wave basis are expanded with a kinetic-energy cutoff of 650 eV. For self-consistent calculations with
plane wave basis the k-point meshes are chosen to be 6$\times$6$\times$6 for the cubic structure, \textcolor{black}{6$\times$6$\times$6 for the triclinic structure,} and 6$\times$6$\times$4 for the monoclinic structure. \textcolor{black}{The convergence
  of the results in terms of choices of k-point mesh is checked by repeating calculations with 12$\times$12$\times$12 (cubic),
  12$\times$12$\times$12 (triclinic) and 12$\times$12$\times$8 (monoclinic) k-points.
  The calculations with larger k-points are found to effect the results only
  marginally, with maximum change in magnetic moment by 0.001 $\mu_B$ and energy difference by 0.75 meV.}
Calculations are carried out with chosen exchange-correlation functional of generalized gradient approximation (GGA)\cite{gga} within the framework of
Perdew-Burke-Ernzerhof (PBE).\cite{pbe} The effect of SOC is included within GGA+SOC implementation of VASP. The effect of missing correlation effect
beyond GGA at Os site is considered within GGA+SOC+$U$ framework\cite{dudarev} by varying the screened Hubbard $U$ value within 2-4 eV and fixing the Hund’s coupling
$J_H$ at 0.6 eV, \textcolor{black}{with choice of fully localized limit (FLL) of double-counting, as implemented in VASP}. \textcolor{black}{The importance of Hubbard $U$ parameter in description of properties of insulating {\it 5d} transition metal oxides have been stressed in a number of literature, especially in context of iridates for which description of its insulating behavior relies on formation of Mott insulating state within the spin-orbit coupled $j=1/2$ manifold.\cite{sriro} Estimated value of Hubbard $U$ in Os oxides and compounds vary from about 2 eV\cite{jac, prb1} to about 3.5 eV.\cite{prb2}
We thus choose $U$ value to vary within the range 2-4 eV.} 
Optimization calculations are done until all the forces on the atoms become smaller than $10^{-5}$ eV/\AA.

The N-th order muffin-tin orbital (MTO) method, \cite{nmto} which relies on self-consistent potential generated by the linear MTO (LMTO) \cite{lmto} method,
is used for constructing effective Os $t_{2g}$ Wannier functions. The low-energy tight-binding Hamiltonian, defined in the effective Os Wannier basis provides
the information of crystal-field splitting at Os site as well as effective hopping interactions between the Os sites. The latter is used as an input to exact diagonalization calculation of the two-site Os problem. The muffin-tin radii of the different atomic sites used in LMTO calculations are chosen as following, \textcolor{black}{2.36 $\AA$ (2.05/1.90 $\AA$) for Sr$_A$, 1.71 $\AA$ (1.99/1.80 $\AA$) for Sr$_B$, 1.25 $\AA$ (1.44/1.39 $\AA$) for Os,
and 0.87 $\AA$ (0.81/0.93 $\AA$, 0.90/0.93 $\AA$, 0.84/0.93 $\AA$) for O atoms in the cubic (monoclinic/triclinic) structure of Sr$_3$OsO$_6$.} 

\section{Results}

\begin{figure}
	\includegraphics[width=0.9\linewidth]{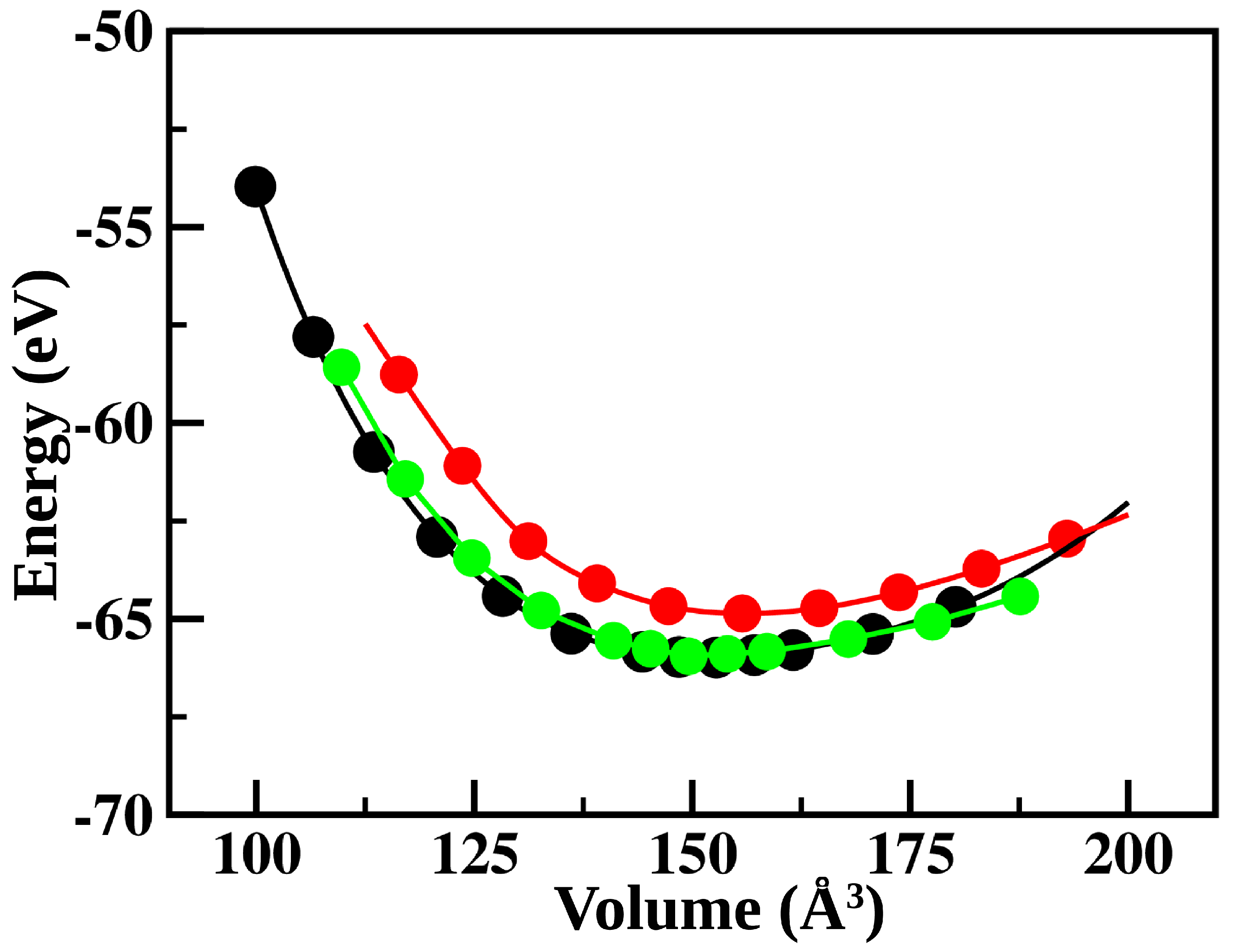}
	\caption{(Color online) Comparison of total energy versus volume calculated within GGA+SOC+$U$ for Sr$_3$OsO$_6$
in cubic (red/dark grey), monoclinic (black) and triclinic (green/light grey) structures.}
\end{figure}

\begin{figure*}
	\includegraphics[width=0.9\linewidth]{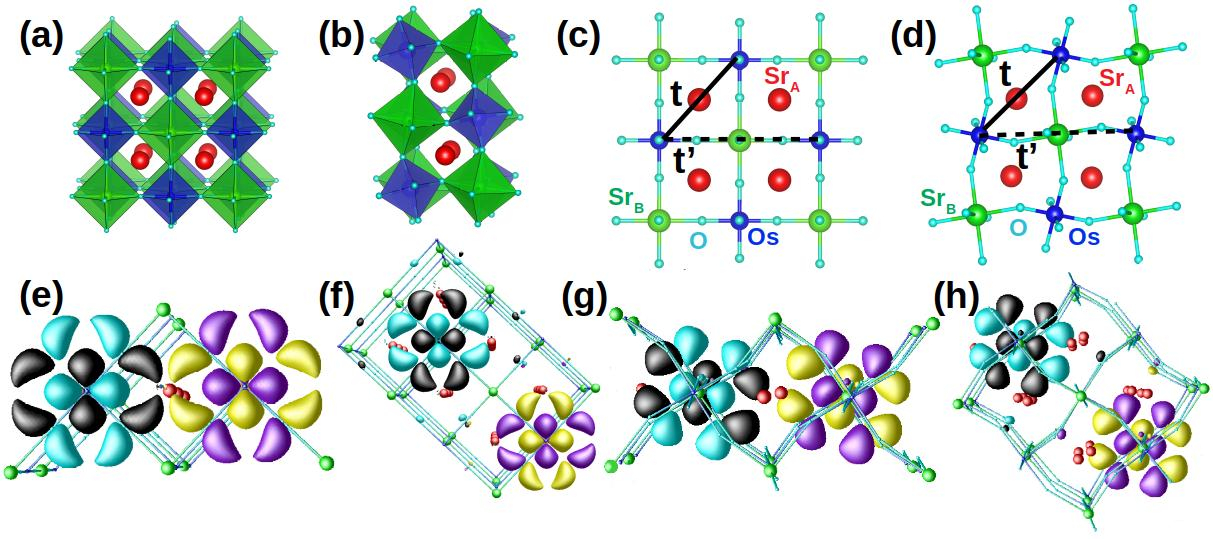}
	\caption{(Color online) (a)-(b) The cubic and distorted monoclinic structures of Sr$_3$OsO$_6$. The Sr atoms
          at A-sites (Sr$_A$) are marked as red (dark grey) balls, the Sr$_B$O$_6$ and OsO$_6$ octahedra at B and B$^{'}$ sites are
          marked in blue (dark grey) and green (light grey), respectively. (c)-(d) The nearest neighbor ($t$) and next-nearest neighbor ($t^{'}$) Os-Os hopping paths across the face and edge
          of the cube and distorted cube in cubic and monoclinic structures. (e)-(f) The overlap of Os effective
          $t_{2g}$ Wannier functions placed at nearest neighbor and next nearest neighbor Os sites.
          Oppositely signed lobes of the Wannier functions are colored differently (cyan (light grey))/black for site 1
          and magenta (dark grey)/yellow (white) for site 2). (g)-(h)
          Same as in (e)-(f) but shown for monoclinic structure.}
\end{figure*} 

\subsection{Crystal Structure of Sr$_3$OsO$_6$}
The ideal space group of rock salt ordered double perovskite is cubic Fm-3m, which is the crystal
structure reported\cite{Os-nat} for the Sr$_3$OsO$_6$ exhibiting high T$_c$ ferromagnetic behavior.
Tilting and rotation of BO$_6$ and B$^{'}$O$_6$ octahedra leads to different lower symmetry
non-cubic structures, tilt and rotation being governed by the tolerance factor, defined as
$t_R = \frac{r_A + r_O}{\sqrt{2}(\frac{r_B + r_B^{'}}{2} + r_O)}$ where $r_A$, $r_B$, $r_B^{'}$
and $r_O$ denote the ionic radii of A site cation, B, B$^{'}$ cations, and O anion, respectively. 
Several different non-cubic structures have been reported for double perovskites, rhombohedral R-3,
tetragonal I4/m and tetragonal I4/mmm, monoclinic P2$_{1}$/n, monoclinic C2/m, with rare
examples of tetragonal P4/mn and triclinic P-1.\cite{Vasala} Compounds with relatively smaller tolerance values,
about 0.92 or less, are predominantly of monoclinic symmetry, while compounds with relatively larger tolerance
factors, beyond 0.92 tend to form in either cubic, or tetragonal or rhombohedral symmetry. Considering the ionic
radii of Sr$^{2+}$, Os$^{6+}$ and O$^{2-}$, the tolerance factor of Sr$_3$OsO$_6$ turns out to be 0.89, similar to that
of Ca$_3$OsO$_6$. Therefore, there is high possibility that Sr$_3$OsO$_6$ grown under solid state reaction would form
in non-cubic structure. In order to resolve this issue, we resort to genetic algorithm as implemented in
USPEX (Universal Structure Predictor: Evolutionary Xtallography)\cite{uspex} which has been demonstrated to predict
accurate crystal structure of different multinary compounds, including double perovskites.\cite{anita1,anita2} We apply the genetic algorithm based
on our first-principles computed energies as goodness parameter, and comparing energies among a large number
of competitive structures over a number of generations. Different structures in each generation are produced
following different variational operations like heredity, mutation, permutation and those generated randomly.
Application of genetic algorithm on Sr$_3$OsO$_6$ results in  monoclinic P2$_{1}$/n and triclinic P-1 as probable
structures for Sr$_3$OsO$_6$. We note that recent high pressure synthesis of Sr$_3$OsO$_6$ in powder
form\cite{Os-powder} did suggest monoclinic and triclinic structures as probable structures for Sr$_3$OsO$_6$. 

In order to rigorously check this issue, we carry out total energy versus volume calculation within GGA+SOC+$U$ scheme
considering the genetic algorithm screened structures, {\it i.e.} monoclinic P2$_{1}$/n and triclinic P-1, and the
reported\cite{Os-nat} cubic structure of Sr$_3$OsO$_6$ thin film. The results are
presented in Fig. 1. Fig. 1 reveals that the energetics of monoclinic and triclinic phases
are close with the energy of monoclinic phase at equilibrium volume being about 30 meV/f.u lower
than that corresponding to triclinic phase. The cubic structure of Sr$_3$OsO$_6$ reported in its
thin film form\cite{Os-nat} is, however, found to be off by a large energy difference of about 1.1 eV/f.u.
from the lowest energy monoclinic structure. The crossover between lowest energy monoclinic and cubic
symmetry would require a large tensile strain of about 6.5$\%$ which appears impractical to achieve.
This is contrary to that reported in Ref. \onlinecite{Os-nat}. This prompts us to conclude that the reported
stabilization of Sr$_3$OsO$_6$ in cubic phase is presumably caused by the epitaxial growth of Sr$_3$OsO$_6$
in a molecular beam epitaxy set up.

In the following we explore the magnetic properties of Sr$_3$OsO$_6$ considering the lowest energy
monoclinic structure and the reported cubic structure in Ref.\onlinecite{Os-nat}, in order to unravel
the role of crystal structure.

\subsection{NMTO-downfolding and Super-exchange paths}

The 6+ nominal valence of Os ion in Sr$_3$OsO$_6$ results in $d^2$ occupancy. The octahedral crystal field of the
surrounding oxygen environment splits the Os 5$d$ levels into three $t_{2g}$'s and two $e_g$'s. Out of three
$t_{2g}$ levels thus two of the levels become half-filled, third one being empty. The empty $e_g$ levels
remain separated from $t_{2g}$'s by large energy splitting of $\approx$ 3-4 eV, suppressing hopping processes
involving $e_g$'s.

This in turn drives antiferromagnetic super-exchange coupling due to virtual hopping between half-filled
$t_{2g}$ orbitals at neighboring Os sites, and ferromagnetic super-exchange coupling due to virtual hopping
between half-filled and empty $t_{2g}$ orbitals at two Os sites. The resultant exchange is decided by the
competition between the antiferromagnetic and ferromagnetic exchanges. For quantitative estimates of the
relative strengths of these two competing interactions, a low energy Os $t_{2g}$ Hamiltonian needs to be
constructed. In order to have a realistic description of this Hamiltonian which incorporates the structural
and chemical information correctly, we employ the NMTO-downfolding.\cite{nmto} Following this procedure,
the low energy tight-binding Hamiltonian is defined in effective Os $t_{2g}$ Wannier basis, obtained by
downfolding or integrating out all the degrees of freedom including Sr, Os $e_{g}$, O $p$ other than
Os $t_{2g}$ degrees of freedom. This results in Os $t_{2g}$ effective Wannier functions with their
head part shaped according to $t_{2g}$ symmetry and tail part shaped according to integrated out orbitals.
With a goal to uncover the influence of underlying crystal symmetry of Sr$_3$OsO$_6$ on magnetic properties,
we construct the low energy Os $t_{2g}$ Hamiltonian in effective Os $t_{2g}$ Wannier basis for cubic symmetry
of Fm-3m, as reported in Ref.\onlinecite{Os-nat} and theoretically predicted lowest energy monoclinic
P2$_1$/n symmetry. The cubic and monoclinic structures are shown in panels (a) and (b) of Fig. 2 which features
corner shared network of OsO$_{6}$ and Sr$_B$O$_{6}$ octahedra with Sr$_A$ atoms occupying the void created by
neighbouring OsO$_{6}$ and SrO$_{6}$ octahedra. Compared to ideal cubic structure, in monoclinic $P2_1/n$
structure the BO$_{6}$ and B$^{'}$O$_{6}$ octahedra exhibit out of plane tilt and in-plane rotation.
The face-centered cubic (FCC) lattice formed by Os ions in cubic structure thus distorts in the
monoclinic structure with Os-O-Os angle across the face of the cube deviating from 90$^{0}$ and
Os-O-Sr angle along the edge of the cube deviating from 180$^{0}$. The deviations of these two angles
in monoclinic phase are found to be 2-5$^{o}$, and 30-40$^{o}$ respectively.

Within the above structural framework, two possible Os-Os hopping paths are, a) the nearest-neighbour
path ($t$) across the face of cube/distorted cube and b) the next nearest neighbour path ($t^{'}$) across
the edge of the cube/distorted cube as shown in Figs 1(c) and (d) for the cubic/monoclinic structures. Following
the conventional wisdom, the $t$ hopping that proceeds through 90$^{o}$ or near 90$^{o}$ Os-O-Os is expected to
be either zero or very weak. The hopping $t^{'}$ proceeding through linear or near linear exchange path
of Os-O-Sr$_B$-O-Os mediated through non magnetic Sr is also expected to be weak. This makes high T$_{c}$
ferromagnetism of Sr$_{3}$OsO$_{6}$ a puzzle.

Surprisingly, the overlap plots of NMTO-downfolding derived Os $t_{2g}$ effective Wannier functions placed
at nearest-neighbour Os sites reveal a remarkable trend. As is evident from \textcolor{black}{Fig. 2(e)}, in the ideal cubic structure,
the O $p$ like tails of the Os $t_{2g}$ Wannier functions strongly bend towards the Sr$_A$ atom at A-site position,
forming a well defined connected path between two Os sites across the face of the cube, mediating the $t$
hopping. Moving to the monoclinic structure, the Wannier functions get misaligned due to the structural
deviation from 90$^o$ (cf Fig. 2(g)), weakening the  connected path between Os sites across the face.
Due to this bending of the O-$p$ like tail, on the other hand, the Wannier functions
have hardly any overlap between two Os sites connected by $t^{'}$ hopping, separated by O-Sr$_B$-O across
the edge (cf Figs. 2(f) and 2(h)).

The trend observed in Wannier function overlap plots gets reflected in the 3 $\times$ 3 tight-binding
Hamiltonian constructed in Os $t_{2g}$ NMTO-downfolding Wannier basis. The monoclinic distortion results
in non-cubic crystal field splitting, lifting the degeneracy of $t_{2g}$ levels, the non-cubic crystal field
splitting being in the range of 0.07-0.10 eV. The nearest neighbour hopping interaction, $t$ which is a
3 $\times$ 3 matrix, following the expectation from Wannier function overlap plots, is found to be significant
with largest strength of 0.17 eV for cubic Sr$_3$OsO$_6$. In the monoclinic phase, this hopping gets largely
suppressed with largest strength of 0.05 eV. Compared to $t$, $t^{'}$ which is also a 3 $\times$ 3
matrix is found to be significantly smaller, the largest strength of $t^{'}$ being about 0.01 eV for the
cubic phase, an order of magnitude smaller compared to that of $t$. For the monoclinic structure, this
is found to be negligibly small. In the exact diagonalization calculations, presented in the following section,
only the effect of hopping $t$ is thus considered, ignoring the small hopping, $t^{'}$.

\begin{figure}
	\includegraphics[width=0.9\linewidth]{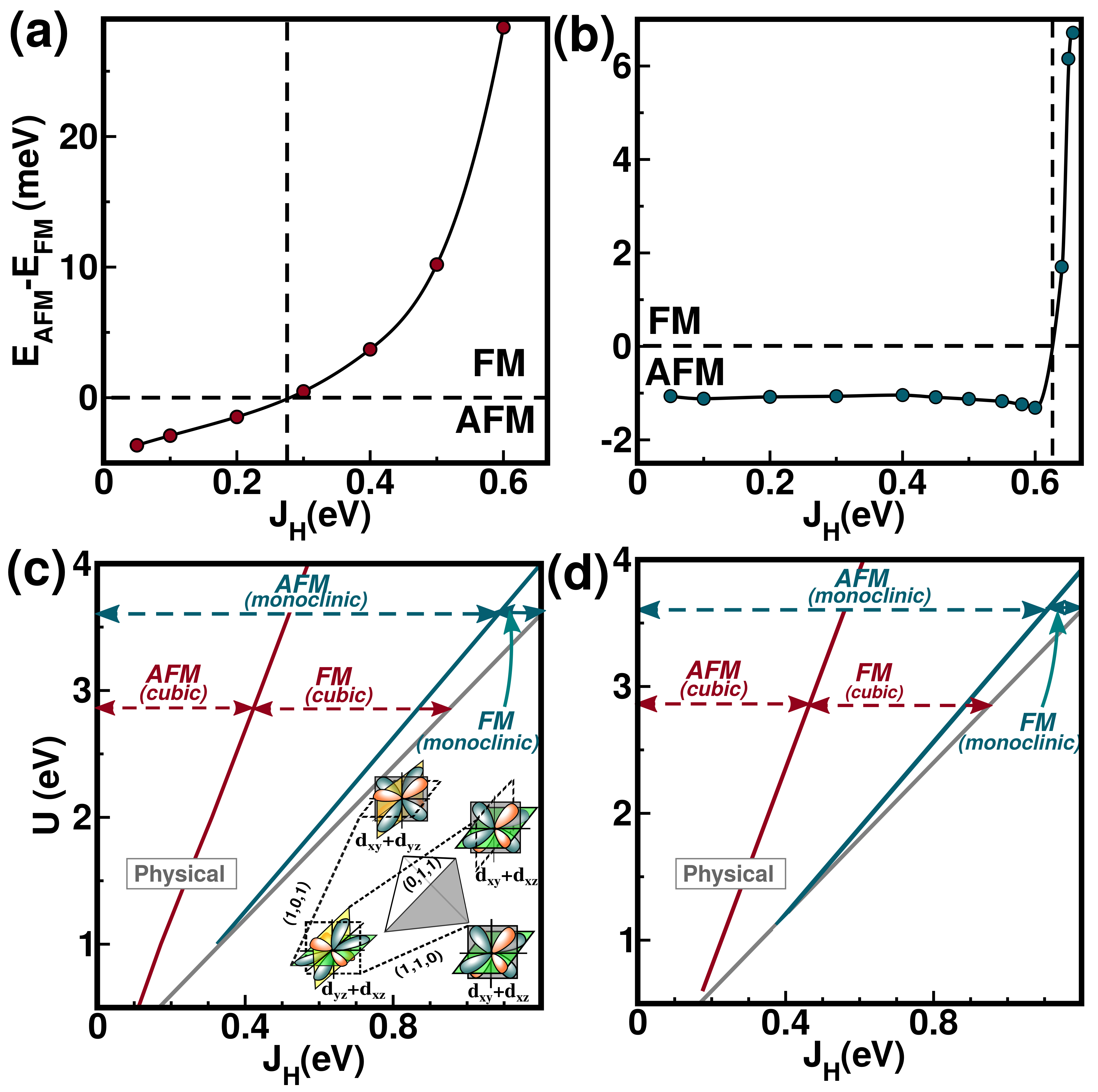}
	\caption{(Color online) The energy difference between ferro and antiferro alignment with Os spins as a function of varying
          values of Hund's coupling, $J_H$, as given in exact diagonalization solution of a two-site Os $t_{2g}$ full multiplet problem, fixing $U$ at 2 eV and $\lambda$ at 0.1 eV and considering the DFT derived hopping interaction in cubic symmetry (panel (a)) and in monoclinic symmetry (panel (b)) of Sr$_3$OsO$_6$. (c) The $U$-$J_H$ phase diagram showing the stabilization of ferro and antiferro alignment of Os spins in cubic and monoclinic phase, for choice of $\lambda$ = 0.2 eV. The inset shows the ordering of
          Os $t_{2g}$ orbitals, with occupancy of two of the Os$^{6+}$ $d$ electrons reversing between $d_{xy}$/$d_{xz}$,
          $d_{yz}$/$d_{xz}$, $d_{xy}$/$d_{yz}$ and $d_{xy}$/$d_{xz}$ along the connecting vectors of Os FCC tetrahedra.
        (d) $U$-$J_H$ phase diagram for choice of $\lambda$ = 0.4 eV. \textcolor{black}{The grey solid line in (c) and (d) demarcates the physically relevant space, $U^{'} = U - 2J_{H} > J_H$.}}
\end{figure}

\subsection{Two-site Model and Magnetic Phase Diagram}

Armed with the knowledge of DFT derived Wannier function overlaps and the low-energy tight-binding Hamiltonian
defined in the Wannier basis, we next explore the interplay between hopping interaction ($t$), crystal field
splitting ($\Delta$), Coulomb interaction ($U$), Hund's exchange ($J_{H}$) and spin-orbit coupling ($\lambda$)
for the cubic and monoclinic structures by performing many-body multiplet calculations within a two-site
problem. This enables us with the opportunity to explore the parameter space in an accurate manner.

The two-site Os $t_{2g}$ only model Hamiltonian taking into account the effect of hopping, Coulomb interaction, and spin-orbit coupling is given by,\cite{ref-1, ref-2, ref-3}

\begin{align}
         & H = H_{\mathrm{onsite}} + H_{\mathrm{int}} +H_{\mathrm{SO}} + H_{t} \nonumber \\
         & H_{\mathrm{onsite}} = \sum_i\sum_{l,m} \Delta^{i}_{~l,m} d^\dagger_{i,l\sigma} d_{i,m\sigma} \nonumber\\
         &H_{\mathrm{int}}= U \sum_i\sum_{l} n_{i,l\uparrow} n_{i,l\downarrow}
         +\frac{(U^{'} -J_{H})}{2} \sum_i\sum_{\substack {l,m \\ l \ne m}} n_{i,l\sigma} n_{i,m\sigma}\nonumber\\
         &+ \frac{U^{'}}{2} \sum_i\sum_{\substack{l,m\\  l\ne m \\ \sigma \ne \sigma^{'}}} n_{i,l\sigma} n_{i,m\sigma^{'}}
         -\frac{J_{H}}{2} \sum_i\sum_{\substack{l,m \\ l\ne m}} (d^{\dagger}_{i,m\uparrow} d_{i,m\downarrow} d^{\dagger}_{i,l\downarrow} d_{i,l\uparrow}\nonumber\\
& + d^{\dagger}_{i,m\uparrow} d^{\dagger}_{i,m\downarrow} d_{i,l\uparrow} d_{i,l\downarrow} +h.c.) \nonumber\\
 &H_{\mathrm{SO}}=\frac{{\mathrm{i}} \lambda}{2}\sum_i \sum_{\substack {l,m,n\\ \sigma,\sigma^{'}}}\epsilon_{lmn} d^\dagger_{i,l\sigma} d_{i,m\sigma^{'}} \sigma^{n}_{\sigma,\sigma^{'}} \nonumber\\
&H_{\mathrm{t}} = \sum_{\substack {l,m,\sigma \\ <i,j>}} t^{ij}_{lm} (d^{\dagger}_{i,l\sigma} d_{j,m\sigma}+h.c.) \nonumber
\end{align}

 $d_{i,l\sigma}$($d^{\dagger}_{i,l\sigma}$) is the annihilation (creation) operator of the $l^{\mathrm{th}}$ orbital ($l =$ 1-3 $\in$ $t_{2g}$) with a spin  $\sigma$ at site $i$ and $n_{i,l\sigma} = d^{\dagger}_{i,l\sigma}d_{i,l\sigma}$.
 $\lambda$ implies the strength of spin-orbit coupling in $H_{\mathrm{SO}}$. 
 $t^{ij}_{lm}$ is the nearest-neighbor hopping between $l^{th}$ orbital of site $i$ and $m^{th}$ orbital of site $j$.
 $H_{\mathrm{onsite}}$ represents the on-site energies of the Os $t_{2g}$ orbitals.  Pure cubic
environment of Os leads to degenerate $t_{2g}$ levels for the cubic structure while all $t_{2g}$ levels
are split due to non-cubic distortion in monoclinic phase. $H_{\mathrm{int}}$ in the Kanamori representation
contains intra-orbital coulomb correlation ($U$), Hund's coupling strength ($J_{H}$), inter-orbital coulomb
correlation ($U^{'}$). The first term of $H_{\mathrm{int}}$ costs an energy $U$ to stabilize the state
counteracting intra-orbital electron-electron repulsion. Hund's coupling ($J_{H}$) in the second term lowers
the energy of states having same spin in multiple orbitals. The inter-orbital Coulomb interaction satisfies the
relation $U^{'}=U-2J_{H}$. The realistic values of 
 non-cubic crystal field and the inter-site hopping parameters are obtained from NMTO\cite{nmto} downfolding 
 calculations, described above, while $U$ and $J_H$ parameters are varied over physically meaningful range
 ($U' > J_H$) with $\lambda$ values fixed at 0.1 eV, 0.2 eV and 0.4 eV. Energy difference between parallel
 and anti-parallel alignment of Os spins is then studied by exact diagonalization of the above two site
 problem.\cite{ref-5} The results are summarized in Fig. 3. The energy differences with a choice of
 $U$ = 2 eV and $\lambda$ = 0.1 eV upon varying strength of $J_H$ are plotted in Figs 3(a) and 3(b) for cubic and
 monoclinic structures respectively. This brings out an important observation. While for the cubic phase, the parallel alignment of Os spin or ferromagnetic interaction becomes favoured over antiferromagnetic interaction
 beyond a $J_H$ value of 0.27 eV or so, for the monoclinic structure  the ferromagnetic interaction gets
 stabilized over antiferromagnetic interaction only beyond $J_H$ value of 0.6 eV. For a choice of
 $J_H$ = 0.6 eV, a reasonable estimate for $5d$ TM like Os, one would thus find the FM interaction is
 stabilized for cubic and AFM interaction is stabilized for monoclinic structure. Further to this, while
 the ferro interaction is stabilized over antiferro in the cubic phase by a large energy gain of almost
 about 30 meV, giving rise to a mean field temperature scale, $\frac{zJs^2}{3}$ ($z$ being number of
 nearest-neighbor which is 12) above 1300 K, the corresponding stabilization of AFM interaction over
 FM interaction in the monoclinic phase is only about 1.5 meV, translating to a mean field temperature
 scale of only $\sim$70 K. One would thus expect ferromagnetic state to support a large T$_c$,  and
 antiferromagnetic state to exhibit a low transition temperature -- a trend found between the reported cubic
 structured Sr$_3$OsO$_6$\cite{Os-nat} and monoclinic structured Ca$_3$OsO$_6$.\cite{ca3oso6} A very
 similar situation is obtained for higher values of $\lambda$ = 0.2 eV and 0.4 eV. The resultant
 $U-J_H$ phase diagram for $\lambda$ = 0.2 eV and 0.4 eV are shown in Figs 3(c) and 3(d) respectively.
 The two phase diagrams appear qualitatively same, except the increase of $\lambda$ value shifts the
 transition from antiferro to ferro to slightly larger value of $J_H$, as spin-orbit coupling tend to
 favour antiferromagnetism. We find that in both choices of $\lambda$ values, the ferromagnetic
 interaction gets stabilized only in a tiny part of the $U - J_H$ phase space in monoclinic structure,
 while a significant part of $U - J_H$ phase space supports ferromagnetic interaction in cubic structure.
 Summarizing these observations, one would thus expect that ferromagnetic phase to be comfortably
 stabilized in cubic structure of Sr$_3$OsO$_6$ with a rather high temperature scale, while the monoclinic
 structure would mostly favour antiferromagnetic interaction with a low transition temperature. The
 antiferromagnetism being frustrated within the FCC motif of the Os sublattice it may also give rise
 to spin-glass like behaviour in compound specific cases. Repeating the calculation considering
 triclinic symmetry of Sr$_3$OsO$_6$, one of the suggested structure of bulk Sr$_{3}$OsO$_6$ in
 Ref.\onlinecite{Os-powder} and the next
 favored theoretically suggested structure, the $U - J_H$ phase diagram for all choices of $\lambda$
 is found to consist of solely antiferromagnetic phase, implying ferromagnetic interaction to be disfavored
 even more in the triclinic phase, compared to the monoclinic phase. This is in complete agreement with
 the observation of antiferromagnetism rather than ferromagnetism in bulk Sr$_3$OsO$_6$.\cite{Os-powder}

 Since the ferromagnetic interaction is favoured by hopping between occupied and empty orbitals the
 stabilized ferromagnetic phase also shows orbital ordering of two occupied orbitals between the
 four sites of FCC tetrahedra, as shown schematically in inset of Fig. 3(c). The two sites connected
 through (1,1,0) vector are found to have the electron occupancy mostly at
 $d_{xz}$/$d_{yz}$ and $d_{xz}$/$d_{xy}$ orbitals, promoting hopping between filled and
 empty $d_{xy}$ orbitals. Similarly orbital occupancy's at sites connected through (1,0,1) and (0,1,1)
 vectors favor hopping between filled to empty $d_{xz}$ and $d_{yz}$ orbitals, respectively.

\begin{figure*}
	\includegraphics[width=0.8\linewidth]{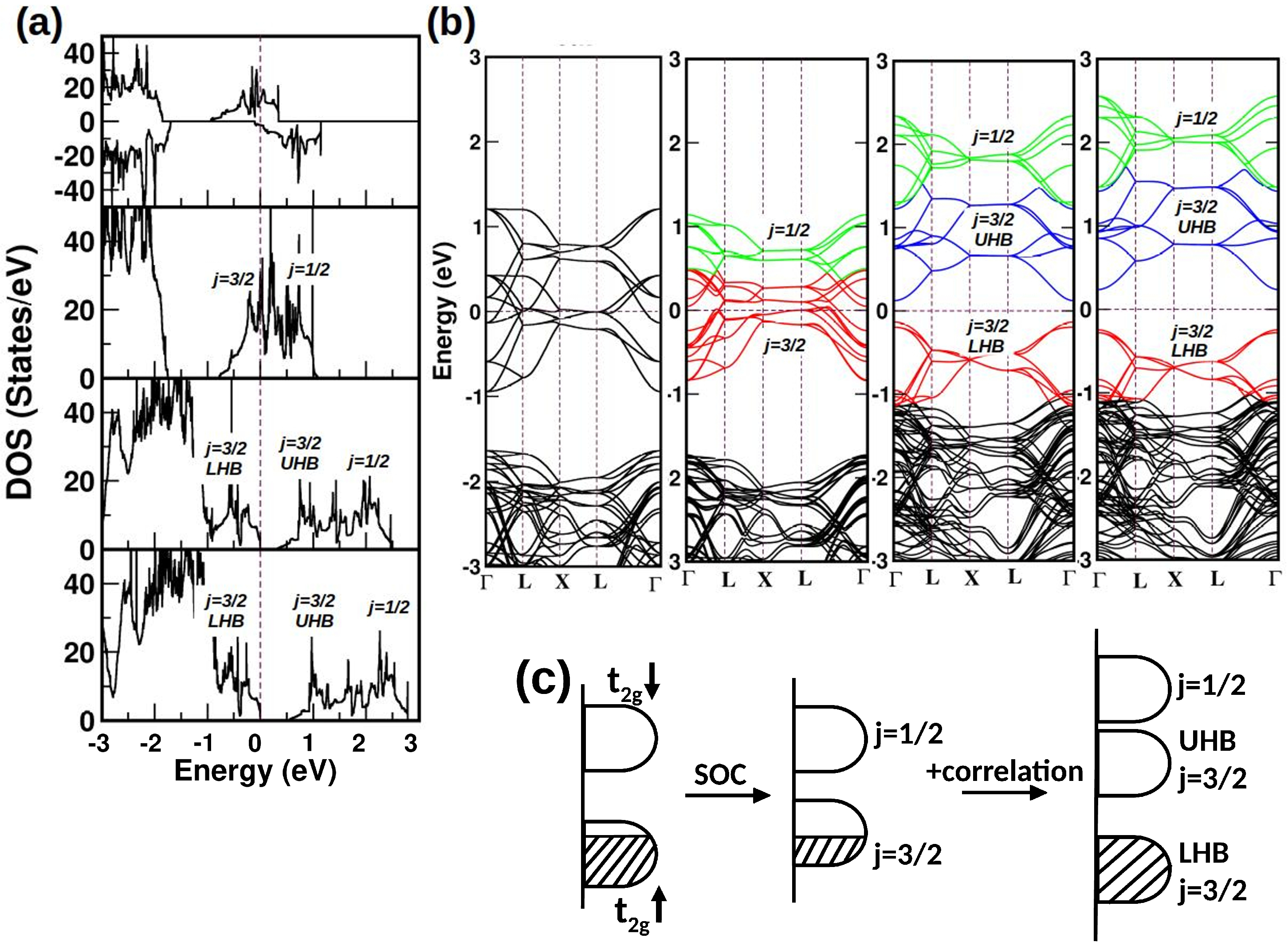}
	\caption{(Color online) (a) The density states of cubic Sr$_3$OsO$_6$ computed within spin-polarized GGA,
          GGA+SOC, GGA+SOC+$U$ [$U$= 2 eV] and GGA+SOC+$U$ [$U$= 3 eV] (from top to bottom). Marked are the spin-orbit
          coupled $j=3/2$ and $j=1/2$ manifolds, which in presence of correlation effect $U$ develops into $j=3/2$ LHB
          and $j=3/2$ UHB. See text for details. (b) The band structures of cubic Sr$_3$OsO$_6$ plotted along high
          symmetry points of the FCC BZ, within GGA, GGA+SOC, GGA+SOC+$U$ [$U$= 2 eV] 
          and GGA+SOC+$U$ [$U$= 3 eV] (from left to right). For spin-polarized GGA band structure, the bands
          corresponding to majority spin channel is only shown. The bands corresponding to 
          $j=3/2$ (LHB, UHB) and $j=1/2$ are colored differently. The zero of the energy in both density of states and
          band structure plots is fixed at respective Fermi level. (c) The systematic effect of turning on SOC and $U$ on
          Os $t_{2g}$s in d$^{2}$ configuration, shown schematically. \textcolor{black}{The filled states are represented as hatched.}
        }
\end{figure*}

\subsection{Insulating Electronic Structure}

Having unraveled the stabilization of ferromagnetic state in cubic Sr$_3$OsO$_6$ being driven by the
large Os-Os nearest neighbor hopping mediated by bending of O $p$ tails of Os $t_{2g}$ Wannier functions
towards the Sr$_A$ atoms at A sites, we next explore the origin of insulating behavior of ferromagnetic,
cubic Sr$_3$OsO$_6$. In order to disentangle various influencing factors in driving the insulating state,
in steps we carry out calculations within the framework of GGA, GGA+SOC and GGA+SOC+$U$.

With two electrons occupying three degenerate $t_{2g}$ levels of Os in an ideal cubic environment, GGA
calculation leads to metallic solution (cf top panel of Fig. 4(a)) with a magnetic moment of 1.10 $\mu_B$
at Os site, and 0.05 $\mu_B$ at O site due to finite Os-O covalency. Inclusion of spin-orbit coupling
within GGA+SOC, develops a significant orbital moment of -0.17 $\mu_B$ at Os site, oppositely aligned to
spin moment of 0.62 $\mu_B$ at Os site, in conformity with less than half-filled nature of Os occupancy.
As is seen from the density of states and band structure plots, presented in Fig. 4(a)-(b), inclusion of SOC,
keeps the solution metallic with Os states crossing the Fermi level. The situation is schematically shown
in panel (c) of Fig. 4. Switching of spin-orbit coupling mixes the up and down spin channels of $t_{2g}$'s,
which effectively behave as $l$=1 quantum number states. The states in presence of spin-orbit coupling
thus are described by four fold degenerate $j=3/2$ and two fold degenerate $j=1/2$ states. With $d^2$
occupancy of Os, the $j=3/2$ states therefore become half-filled, a situation very
similar to $d^5$ Iridate like Sr$_2$IrO$_4$ which results in half-filled $j=1/2$ states
instead of half filled $j=3/2$ states in the present case.\cite{sriro} Akin to Sr$_2$IrO$_4$,
one would thus expect inclusion of onsite correlation, $U$, modeled within a GGA+SOC+$U$ framework would
be able to open up gap. This expectation turned out
to be true where a gap within $j=3/2$ manifold is found to open up for a choice of $U$ value $\ge$ 2 eV.
As shown in Fig. 4(b), within the GGA+SOC+$U$ description,
out of 24 $t_{2g}$ bands of Os in a 4 f.u. cubic cell, 8 of the bands lie below the Fermi level, forming $j=3/2$ lower Hubbard bands (LHB), while rest 16 of the bands lie above Fermi level, separated by a gap from $j=3/2$ LHB states. Eight empty $j=3/2$ upper Hubbard bands (UHB)
overlap with eight empty  $j=1/2$ bands due to finite band widths
of $j=3/2$ and $j=1/2$. With inclusion of Hubbard $U$ correction, both the spin moment and orbital moment
of Os show an increase with values 1.48 $\mu_B$ and -0.65 $\mu_B$ respectively for choice of $U$ = 2 eV.
This leads of a magnetic moment of 0.83 $\mu_B$, in good agreement with experimentally
measured\cite{Os-nat} moment of 0.77 $\mu_B$.
Like a Mott insulator, the gap value is found to scale with $U$, with calculated band gap value of
0.33 eV for $U$ = 2 eV and 0.55 eV for $U$ = 3 eV. Thus Sr$_3$OsO$_6$ forms an example of spin-orbit
entangled Mott state in $j=3/2$ sector, similar to that of Sr$_2$IrO$_4$ in $j=1/2$ sector.

\section{Summary and Discussion}

In this study, we take up the case of Os containing double perovskite compound, Sr$_3$OsO$_6$, and study the microscopic origin of its reported high T$_c$ ferromagnetic insulating behavior.\cite{Os-nat} It is curious to note that while cubic crystal structure of Sr$_3$OsO$_6$ has been reported in Ref.\onlinecite{Os-nat} grown as thin film, a recent report of bulk Sr$_3$OsO$_6$ compound synthesized via solid-state route suggests the crystal structure to be in non-cubic monoclinic or triclinic symmetry.\cite{Os-powder} We thus start our study by exploring the
crystal symmetry of Sr$_3$OsO$_6$ predicted via application of genetic algorithm together with first-principles total energy calculations. These calculations show monoclinic symmetry as most preferred symmetry
followed by triclinic symmetry, while the cubic symmetry structure lie energetically much higher.

Following this finding, we explore the influence of crystal symmetry on the magnetic properties by
considering the predicted lowest energy monoclinic structure and the reported\cite{Os-nat} cubic structure.
Towards this end, employing first-principles derived Wannier representation of Os $t_{2g}$ orbitals, we
first show that Os-Os super-exchange is primarily governed by counter-intuitive large Os-Os hopping across
the face of the cubic structure, which dominates over conventionally expected Os-O-Sr$_B$-O-Os super-exchange path along the edge of the cubic structure. This hopping process, which is caused by Sr-O covalency driven
bending of O $p$ like tails of the Os $t_{2g}$ effective Wannier functions, gets substantially reduced in the
distorted monoclinic phase due to misalignment of O $p$ like tails.
Using DFT derived Os $t_{2g}$ low energy Hamiltonian defined in the Wannier basis, we next solve the two-site Os $t_{2g}$ full multiplet problem within exact diagonalization scheme in $t$-$U$-$J_H$-$\lambda$ space. We find a crossover from antiferro alignment of Os spins to ferro alignment of Os spins happens in $U$-$J_H$ space for several
choices of $\lambda$ values. For the choice of $\lambda$ values relevant for Os, a large part of $U$-$J_H$ phase diagram in cubic symmetry is found to support ferro alignment
of Os spins. On the contrary, ferro phase is found to be largely suppressed in
monoclinic symmetry, the phase space being primarily dominated by antiferro alignment of Os spins.
Interestingly, for choice of same parameter values of $U$, $J_H$ and $\lambda$, the stabilization energy of ferro alignment over antiferro in the cubic phase is found
to be more than order of magnitude larger compared to stabilization energy of antiferro alignment of Os spins over ferro in distorted monoclinic phase. This rationalizes
the observation of high T$_c$ in reported cubic, FM phase of Sr$_3$OsO$_6$ and low Ne\'el temperature in reported monoclinic, AFM phase of Ca$_3$OsO$_6$. This analysis establishes that the stabilization of cubic phase of Sr$_3$OsO$_6$ in Ref.\onlinecite{Os-nat}, was crucial to its observed high T$_c$ ferromagnetism. Following this
understanding, we investigate the origin of insulating behavior of FM Sr$_3$OsO$_6$ through systematic analysis of GGA, GGA+SOC and GGA+SOC+$U$ scheme of electronic
structure calculations. While both GGA and GGA+SOC calculations lead to metallic solutions, insulating ground state is achieved only within the treatment of GGA+SOC+$U$
with reasonable choices of $U$ values. This characterizes the insulating state in cubic Sr$_3$OsO$_6$ as
$j=3/2$ Mott state induced by large spin-orbit coupling at Os site.

Finally, searching for related double perovskite candidates, exhibiting high T$_c$ FM
insulating state, Sr$_2$CaOsO$_6$ appear to be a promising candidate. With Sr at A site, and Ca at B$^{'}$
site, the tolerance factor of this compound turns out to be 0.93. Following the expectation from
the tolerance factor, the powder X-ray diffraction analysis\cite{sr2mgoso62} of synthesized Sr$_2$CaOsO$_6$
suggested an ordered cubic structure. Calculated Os-Os hopping strengths in NMTO-downfolding scheme of
Wannier representation in Sr$_2$CaOsO$_6$ turn out to be rather similar to that of cubic Sr$_3$OsO$_6$,
the largest Os-Os hopping being 0.16 eV compared to 0.17 eV in Sr$_3$OsO$_6$. While there exists report
of synthesis of this compound,\cite{sr2mgoso62} to the best of our knowledge its magnetic property, so far
remains unexplored. This raises the hope that the bulk form of Sr$_2$CaOsO$_6$ should also support high T$_c$ ferromagnetism. In view of our study, it will be worth while to take this up.

\section{Acknowledgement}

T.S-D acknowledges computational support under Thematic Unit of Excellence funded by Department of Science and Technology, India. T.S-D thanks Arun Paramekanti or useful discussions. I.D thanks Science and Engineering Research Board (SERB), India (Project No. EMR/2016/005925) for financial support.

\end{document}